\documentclass[journal]{IEEEtran}
\usepackage{graphicx} 
\usepackage{caption} 
\usepackage{amsmath} 

\begin{document}

\title{LDPC-Coded Molecular Communications with Increased Diversity}

\author{
    \IEEEauthorblockN{Alpar Türkoğlu\IEEEauthorrefmark{1}\IEEEauthorrefmark{2}, Berk Karabacakoğlu\IEEEauthorrefmark{1}, Ali Emre Pusane\IEEEauthorrefmark{1}}\\
    \IEEEauthorblockA{\IEEEauthorrefmark{1}Department of Electrical and Electronics Engineering, Bogazici University, Istanbul, Turkey \\ \IEEEauthorrefmark{2} Department of Physics, Bogazici University, Istanbul, Turkey \\
    Email: \{alpar.turkoglu, berk.karabacakoglu, ali.pusane\}@boun.edu.tr}
}


\maketitle

\begin{abstract}
This paper suggests achieving diversity gains while utilizing low-denisty parity check (LDPC) codes in molecular communications. Intersymbol interference (ISI) causes a significant disadvantage in error performance for molecular communications. Even though decoding LDPC codes with soft decoding yields a considerable enhancement in the bit error rate (BER) curves, this can be further improved by utilizing diversity gain. In order to achieve this, two different messenger molecule types are sent to transmit the message codeword and its interleaved version. The molecular communication channel is then modeled, and the error performance of the proposed method is estimated by Monte-Carlo simulations. This approach provides considerable improvement in the error performance in the scenario where few messenger molecules are transmitted per bit.
\end{abstract}

\begin{IEEEkeywords}
Molecular communications, communication via diffusion, ISI mitigation, LDPC codes, diversity gain. 
\end{IEEEkeywords}

\IEEEpeerreviewmaketitle

\section{Introduction}

\IEEEPARstart{W}{ith} the rapid advancement of technology, novel information transfer techniques are continually emerging. Among these, molecular communication stands out as a pivotal development. Departing from conventional methods that rely on electromagnetic waves, molecular communication harnesses the power of messenger molecules as information carriers, mirroring the intricate communication processes within biological cells. This unique approach positions molecular communication as a promising solution with vast potential for diverse applications in the fields of biology, medicine, and environmental science.
 
A basic molecular communication system can be conceptualized as a single input-single output (SISO) system, comprising a point transmitter and a spherical receiver. When information is present, the point transmitter releases $N$ molecules, and it stays silent when the opposite condition occurs. These released molecules undergo Brownian motion to reach the receiver. However, the inherent stochastic nature of this motion can introduce errors due to random fluctuations in the positions and velocities of molecules. This underscores the critical need for an effective error control mechanism in molecular communication.
 
In the literature, some techniques are introduced to molecular communication to decrease the effects of errors. To begin with, the adaptive threshold detection method \cite{7390096} estimates a threshold for the number of the arriving molecules to the spherical receiver in a given period of time to decide whether the transmitter sent any information. However, this approach does not provide a sufficient solution to avoid intersymbol interference (ISI). ISI is a phenomenon in which the messenger molecules sent due to the information are not received by the receiver in the given period, and may be wandering in the communication channel. This increases the probability that the receiver may receive them in the subsequent period. After a while, as the number of molecules that wander in the channel increases, the receiver may make errors due to ISI. This is a key challenge in molecular communication that error control mechanisms aim to address.
In the method \cite{7118126} pre-equalization, an extra transmitter eliminates the ISI by sending anti-information-carrying molecules when the transmitter sends no information. This approach gives a much better result than the hard decision \cite{7390096} technique for large number of messenger molecules per bit.

Error control codes are techniques used in information and data transmission to detect and correct errors that may occur during the communication process. These codes add redundant information to the original data, allowing the receiver to identify and correct errors caused by noise, interference, or other transmission issues. The primary goal of error control codes is to ensure the transmitted information's accuracy and reliability. One of the most preferred and novel error control codes is low-density parity-check (LDPC) codes \cite{gallager1963ldpc}, which offer high performance for a large amount of data while enabling easy implementation of decoders. 
LDPC codes show high success in traditional communication systems that use electromagnetic waves having symmetric error possibilities. However, molecular communication systems are vulnerable to asymmetric errors. Still, various methods can be developed to implement error control codes in molecular communication. 

While different channel coding methods, including LDPC, have been previously implemented in molecular communications systems \cite{7273857}, the one-step majority logic decoder does not fully utilize the channel model's statistical properties, leading to suboptimal performance. Therefore, it is crucial to explore different decoding techniques, such as belief propagation \cite{10.5555/534975}, a soft decision technique that uses the LLR values for decision-making, to achieve better performance results. In this paper, we propose to utilize LDPC codes with diversity gain in molecular communications.

\section{System Model}

\begin{figure}[h]
    \centering
    \includegraphics[width=0.9\linewidth]{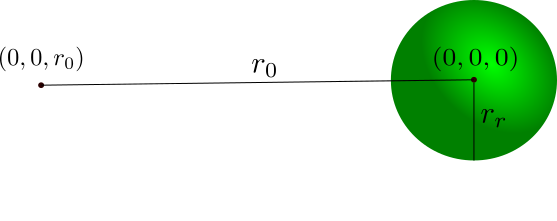} 
    \caption{A simple SISO model}
    \label{fig:block_diagram}
\end{figure}

\subsection{The molecular communication model}
The SISO model assumes a spherical receiver at the origin (0, 0, 0) with a precise radius $r_r$ and a point transmitter, located at (0, 0, $r_0$). The transmitter releases $N$ molecules at the onset of the simulation. These molecules execute a random Brownian motion with a mean of 0 and variance of $2D\Delta t$, where $D$ is the diffusion coefficient, and $\Delta t$ is the sampling period. This simulation is conducted for $T$ seconds. The exact parameters that have been used can be seen in Table 1. Figure 1 offers a brief view of the simulated system.
In Figure 2, the channel response (P) vector values for intervals of size $T_s$ are obtained by integrating the received molecule numbers in each interval and normalizing by dividing the total molecule number $N$. Since running this simulation is a computationally inefficient process, the Gaussian approximation \cite{10.1145/2800795.2800816} for the transmission process given as

\begin{equation}
M_{R_i} = \sum_{l=1}^{L} bit_{i-l} \left[ \sim N \left( M_M P_{l+1}, \sqrt{M_M P_{l+1} (1 - P_{l+1})} \right) \right]
\end{equation}
has been utilized, where $L$ is the channel memory, $M_M$ and $M_R$ represent the sent and received number of molecules, respectively.

\subsection{The LDPC model}
 At the transmitter, $n$-bit LDPC encoded codeword \textbf{c} is obtained by the multiplication of the k-bit codeword message vector \textbf{u} with the $k{\times}n$ generator matrix \textbf{G}. At the receiver, the belief propagation algorithm decodes the message, which primarily relies on the log-likelihood ratio (LLR) values and the parity check matrix \textbf{H}. For this purposes, $3{\times}6$ regular LDPC codes have been used.

\subsection{Combined model}
When applying the BP decoder to molecular communication, the computation of the LLR values is a key task. Past suggestions have leaned towards using Poisson random variable approximation for this \cite{DAMRATH201898}. However, this simulation yields a significant challenge with numerical overflows due to the high number of molecules. To address this, a generalized Gaussian random variable approximation is preferred that effectively mitigates these overflows. This approximation yields
\begin{equation}
LLR_i = \log \frac{
    \sum_{k= 1}^{2^{L-1}} f(MR_i \mid \{\mathbf{c}[i] = 0 \ \& \ \mathbf{c}_k[i-L:i-1]\})
}{
    \sum_{k=1}^{2^{L-1}} f(MR_i \mid \{\mathbf{c}[i] = 1 \ \& \ \mathbf{c}_k[i-L:i-1]\})
}
\end{equation}
where f(x) is a Gaussian PDF with its mean and variance is calculated from the all possible combinations of the previous $L-1$ bits of \textbf{c} as given in (1).

\begin{figure}[h]
    \centering
    \includegraphics[width=0.9\linewidth]{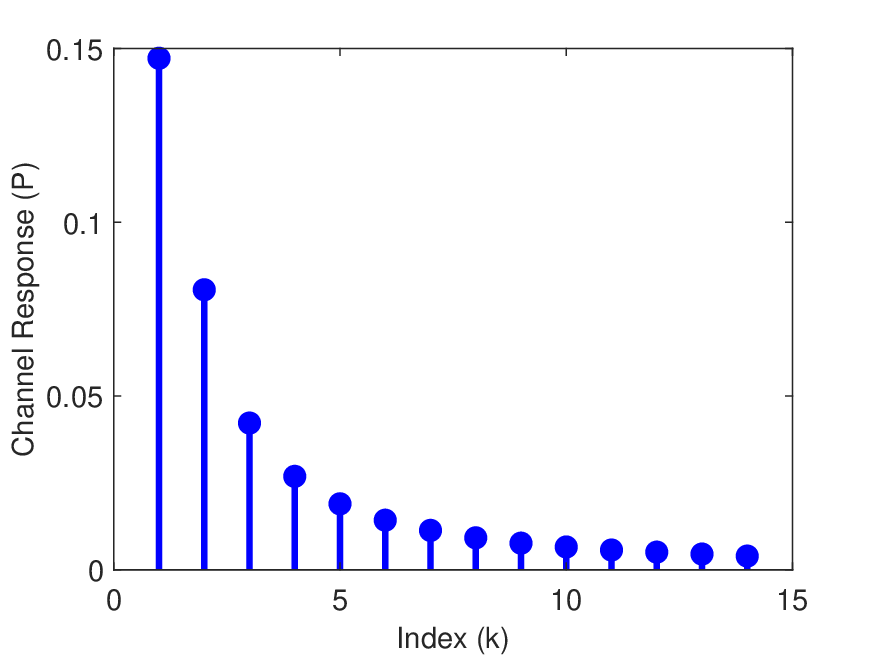} 
    \caption{Channel response for $T_s$ = 150ms}
    \label{fig:block_diagram}
\end{figure}

\begin{table}[h!]
\centering
\caption*{Table 1: Parameters used in molecular communication system}
\resizebox{\linewidth}{!}{%
\begin{tabular}{|c|c|c|}
\hline
\textbf{Name} & \textbf{Description} & \textbf{Value} \\ \hline
T & Time period & 2.1s \\ \hline
D & Diffusion coefficient & 79.4$\mu$$m^2$/s \\ \hline
$r_0$ & Distance between transmitter and receiver & 10$\mu$m \\ \hline
$r_r$ & Radius of receiver & 5$\mu$m\\ \hline
N & Total molecule number & $10^6$ \\ \hline
$\Delta t$ & Sampling period & 0.1ms \\ \hline
\end{tabular}%
}
\label{tab:parameters}
\end{table}

\subsection{Diversity Gain}
Introducing a second molecule type to the communication system to decrease the effects of ISI is tried to get a better BER curve. For this purpose, the diversity gain approach is proposed in this work. Diversity gain, in this context, refers to the improvement in the system's performance due to the use of multiple transmission paths. This method includes transmitting two different codewords through the channel. One is a standard codeword, and the other is an interleaved codeword. When the location of the bits is changed, the ISI effect originated by the previous bits may be altered. These codewords are simultaneously sent to the receiver through the channel; at the receiver, LLR values of the same encoded bit are first calculated and then equally weighted. The aim is that even if the LLR value of one codeword may lead to an erroneous decision, information from the other codeword may help reduce the errors as the equally weighted LLR makes the decision. A simple block diagram for the overall system model can be found in Figure 3.

\begin{figure}[h]
    \centering
    \includegraphics[width=0.9\linewidth]{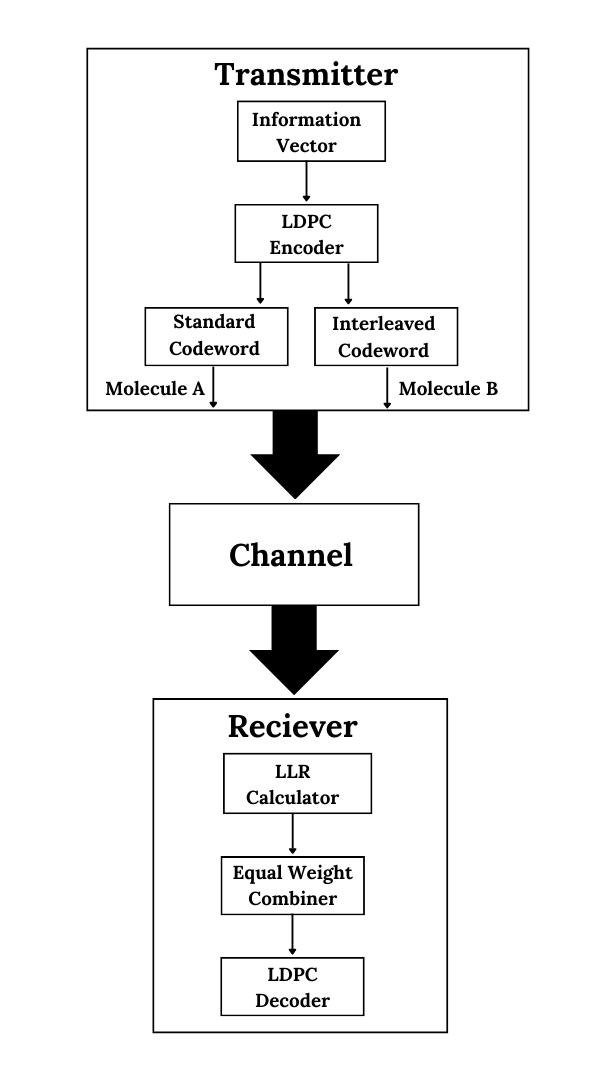} 
    \caption{The Block Diagram for the Proposed System}
    \label{fig:block_diagram}
\end{figure}

\section{Numerical Results and Performance Evaluation}
Monte Carlo simulations have been employed to evaluate the performance of the proposed method. $1{\times}100$ random binary vectors were created and encoded with 3x6 regular LDPC codes. A maximum of 10 iterations of BP decoding were performed for each codeword at the receiver. The simulation runs until 1020 frame errors are counted, providing significant insights into the method's performance. Bit error rate (BER) curves have been plotted against the Number of Molecules for Bit-1 ($M_M$). The SISO parameters and the $P$-vector are given in Table 1 and Figure 2, respectively. The channel memory consists of the last 1.4 seconds for computational concerns. The total amount of molecules used should be equal for both scenarios to obtain a fair comparison. Thus, $M_M$ values for the diversity gain method are taken as the sum of molecule types A and B. 
\par
Figure 4 indicates that the diversity gain method yields a significant decrease in the BER compared to the single molecule case. This stems from the ISI caused by the asymmetry of the channel, which has been overcome by introducing the new information-carrying molecule. Specific repetitions of the bit-1 increase the ISI in this channel type; however, randomly interleaving bits and then averaging the resulting LLRs reduces this effect.
\par
Figure 5 compares the previously discussed pre-equalization method \cite{7118126} with the proposed diversity gain method. Even though pre-equalization yields a clear advantage for a large number of molecules, the diversity gain method still offers a great solution, since it is essential to obtain better error performance in the low molecule region.

\begin{figure}[h]
    \centering
    \includegraphics[width=0.9\linewidth]{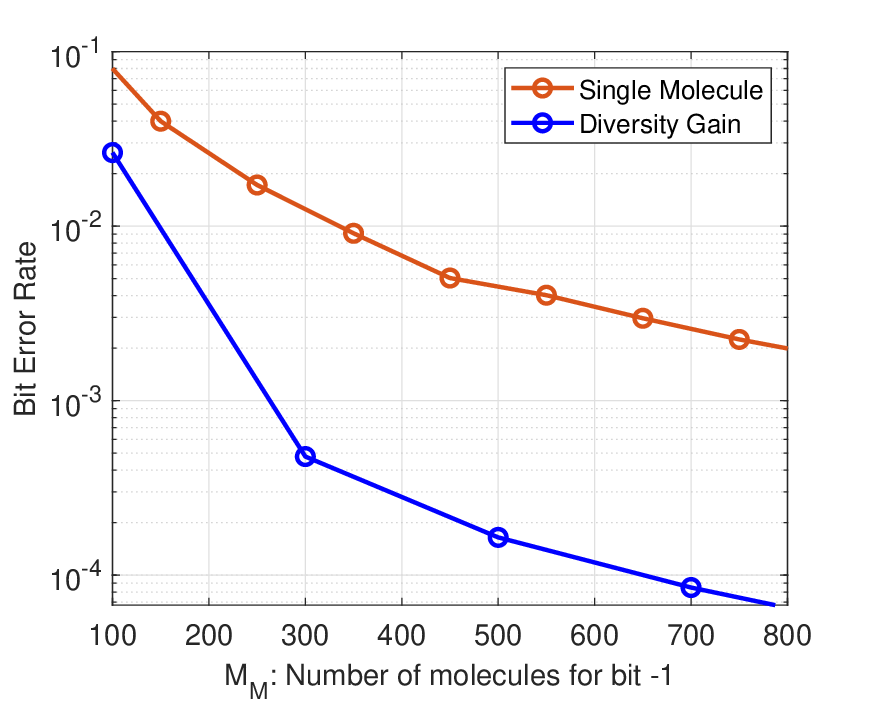} 
    \caption{BER curves of single molecule and diversity gain LDPC codded SISO system $T_s$ = 150ms}
    \label{fig:block_diagram}
\end{figure}

\begin{figure}[h]
    \centering
    \includegraphics[width=0.9\linewidth]{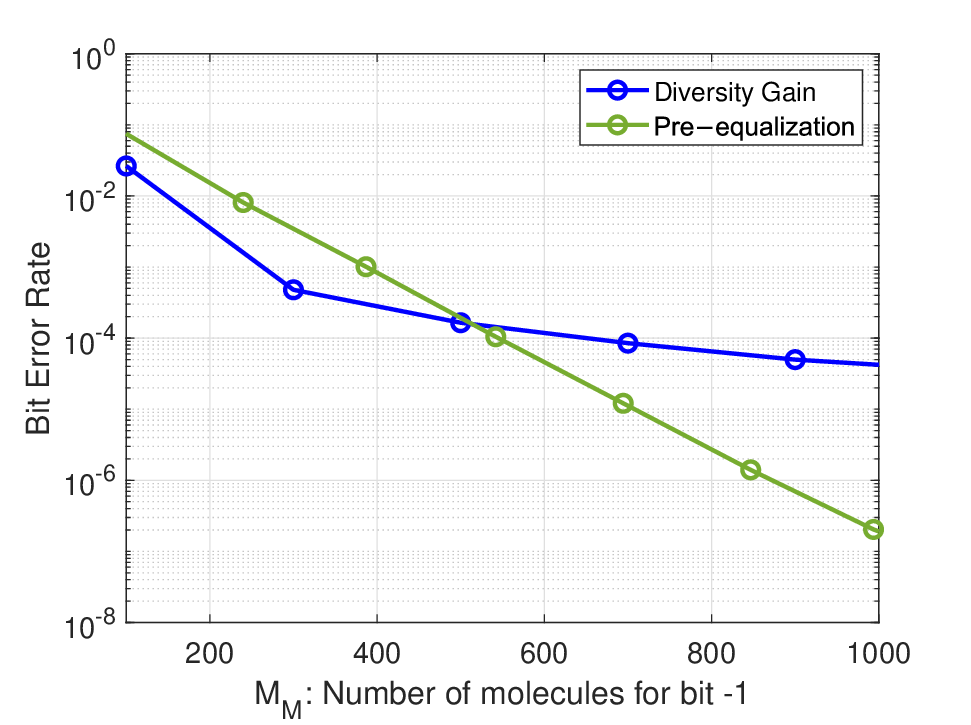} 
    \caption{BER curves of diversity gain LDPC codded and pre-equalization SISO system $T_s$ = 150ms}
    \label{fig:block_diagram}
\end{figure}

\section{Conclusion}
In this paper, LDPC codes are implemented with a diversity gain approach for diffusion-based molecular communication systems. For this purpose, first, the information codeword is encoded using $3{\times}6$ regular LDPC codes. Then, it is transmitted with its interleaved version using two different molecule types through the channel, modeled using a Gaussian approximation. On the receiver side, the LLR values for the incoming molecules of both types were calculated using a similar approximation before taking their average. These averaged LLR values are then used in the soft decoding process of LDPC codes. 
\\
The error performance of this method is evaluated with Monte-Carlo simulations. The introduction of diversity gain is shown to improve the error performance of the LDPC codes for all numbers of molecules. Comparison with the pre-equalization method shows that even though using this method can't offer any advantage for high number of molecules, as moving to the low molecule number regime, the diversity gain method outperforms its alternatives. Since it is more critical to use the molecule resources efficiently without sacrificing much of the error performance, this method offers a reasonable solution for practical molecular communication system.



\bibliographystyle{IEEEtran}
\bibliography{references}

\end{document}